\documentclass[11pt, titlepage]{article}

\usepackage{titlesec} 

\titleformat{\subsection}[runin]
{\normalfont\large\bfseries}{\thesubsection}{1em}{}
\titleformat{\subsubsection}[runin]
{\normalfont\normalsize\bfseries}{\thesubsubsection}{1em}{}

\usepackage{helvet} 

\usepackage[T1]{fontenc}

\usepackage{xr}
\makeatletter

\newcommand\sixteen{\@setfontsize\sixteen{14pt}{6}}
\renewcommand{\maketitle}{\bgroup\setlength{\parindent}{0pt}
	\begin{flushleft}
		\vspace{-.375in}
		\sixteen\bfseries \@title \\
		\medskip
	\end{flushleft}
	\@author \\
	\@date
	\egroup
}
\makeatother

\title{Navigation of brain networks}
\author{Caio Seguin \footnote{Melbourne Neuropsychiatry Centre, Department of Psychiatry, The University of Melbourne and Melbourne Health, Carlton South, Victoria, Australia. Corresponding author: \href{mailto:caioseguin@gmail.com}{caioseguin@gmail.com}.}, Martijn P. van den Heuvel\footnote{Department of Psychiatry, Brain Center Rudolf Magnus, University Medical Center Utrecht, Utrecht, The Netherlands}, Andrew Zalesky\footnote{Melbourne Neuropsychiatry Centre, Department of Psychiatry, The University of Melbourne and Melbourne Health, Carlton South, Victoria, Australia; Melbourne School of Engineering, The University of Melbourne, Parkville, Victoria, Australia}}

\date{January 24, 2018}

\usepackage{amsfonts, amsmath, amsthm, amssymb} 
\usepackage{graphicx} 
\usepackage{booktabs} 
\usepackage{wrapfig} 
\usepackage[labelfont=bf]{caption} 
\usepackage[top=1in,bottom=1in,left=1.25in,right=1.25in]{geometry} 
\usepackage[nocompress]{cite}
\usepackage{caption}
\usepackage{graphicx}
\usepackage{subfig}

\usepackage[usenames, dvipsnames]{color}
\usepackage{hyperref}

\begin{document}

\maketitle

\section*{Abstract}

Understanding the mechanisms of neural communication in large-scale brain networks remains a major goal in neuroscience. We investigated whether navigation is a parsimonious routing model for connectomics. Navigating a network involves progressing to the next node that is closest in distance to a desired destination. We developed a measure to quantify navigation efficiency and found that connectomes in a range of mammalian species (human, mouse and macaque) can be successfully navigated with near-optimal efficiency (>80\% of optimal efficiency for typical connection densities). Rewiring network topology or repositioning network nodes resulted in 45\%--60\% reductions in navigation performance. Specifically, we found that brain networks cannot be progressively rewired (randomized or clusterized) to result in topologies with significantly improved navigation performance. Navigation was also found to: i) promote a resource-efficient distribution of the information traffic load, potentially relieving communication bottlenecks; and, ii) explain significant variation in functional connectivity. Unlike prevalently studied communication strategies in connectomics, navigation does not mandate biologically unrealistic assumptions about global knowledge of network topology. We conclude that the wiring and spatial embedding of brain networks is conducive to effective decentralized communication. Graph-theoretic studies of the connectome should consider measures of network efficiency and centrality that are consistent with decentralized models of neural communication.

\section*{Introduction}

Nervous systems are networks and one of the key functions of a network is to facilitate communication. While much is known about the properties of brain networks across various species and scales, from the neuronal nervous system of the nematode {\it C. Elegans} \cite{varshney:2011}, to the large-scale human connectome \cite{sporns:2005, hagmann:2008}, little is known in systems neuroscience about the network principles that govern neural communication and signalling.   Complex topological properties such as small-worldness \cite{watts:1998, bassett:2006}, modularity \cite{meunier:2010} and a core of highly interconnected hubs \cite{heuvel:2012} are universally found across advanced and simple species, including  mouse \cite{oh:2014, rubinov:2015}, macaque \cite{harriger:2012, markov:2014} and human connectomes \cite{heuvel:2016}. Support for efficient communication between neuronal populations is conjectured to be one of the main adaptive advantages behind the emergence of these complex organizational properties \cite{bullmore:2012, misic:2015}.

Understanding how neural information is routed and communicated through complex networks of white matter pathways remains an open challenge for systems neuroscience \cite{fornito:2016, avena:2017}. To date, connectomics has largely focused on network communication based on optimal routing \cite{bullmore:2009, rubinov:2010}, which proposes that information traverses the shortest path between two nodes. For nodes that are not directly connected, the shortest path is polysynaptic and must traverse one or more intermediate nodes. However, identifying shortest paths requires individual elements of nervous systems to possess global knowledge of network topology. This requirement for centralized knowledge has been challenged on the basis that biological systems are decentralized (i.e., individual system components operate based on local knowledge), motivating recent work on alternative communication strategies to model large-scale neural information transfer  \cite{goni:2013, goni:2014, misic:2015, avena:2016}.

Navigation is one such strategy \cite{kleinberg:2000}. Navigating a network is as simple as progressing to the next node that is closest in distance to a desired target. Navigation is not guaranteed to successfully reach a target destination. Moreover, targets might be reached using long, inefficient paths. However, several real-world networks are known to be efficiently navigable, including biological, social, transportation and technological systems \cite{boguna:2009, gulyas:2015, allard:2017}. Successful navigation depends on certain topological properties such as small-worldness \cite{kleinberg:2000} and a combination of high clustering and heterogeneous degree distribution \cite{boguna:2009}, all of which are found in the brain networks of several species \cite{heuvel:2016}. 

Here, we comprehensively investigate the feasibility of navigation routing as a model for large-scale neural communication. We develop a novel measure of navigation efficiency and apply it to publicly available connectomic data acquired from the macaque, mouse and human brain. We find that brain networks are highly optimized for successful and efficient navigation. Specifically, we show that rewiring connections (topological randomization) or repositioning nodes (spatial randomization) drastically reduces navigation efficiency. We attempt to evolve (clusterize or randomize) brain networks {\it in silico}, with the goal of improving navigability but find that the margin for improvement is minimal. Finally, we characterize the centrality of nodes under navigability and investigate the relation between navigation path lengths and functional connectivity (FC) inferred from resting-state functional magnetic resonance imaging (MRI). Compared to shortest path routing, we find that navigation utilizes the brain's resources more uniformly and yields stronger correlations with FC.

\section*{Results}

\subsection*{Navigation performance measures.}

We consider neural communication from the graph-theoretic standpoint of delineating paths (routes) in the connectome between pairs of nodes (gray matter regions). While neural communication can be modelled as a diffusion process where information is broadcast in parallel over multiple paths \cite{estrada:2008, goni:2013}, here we only consider single-path communication (i.e., routing). A routing strategy defines a set of rules for identifying a path from a source node to a target node. Path length refers to the number of connections that comprise a path (hops) or the sum across the lengths of these connections. To minimize conduction latency, noise introduced by synaptic retransmission and metabolic costs, neural communication should take place along paths with short path lengths \cite{fornito:2016, avena:2017}.

\begin{figure}
	
	\centering
	\includegraphics[width=0.6\linewidth,keepaspectratio]{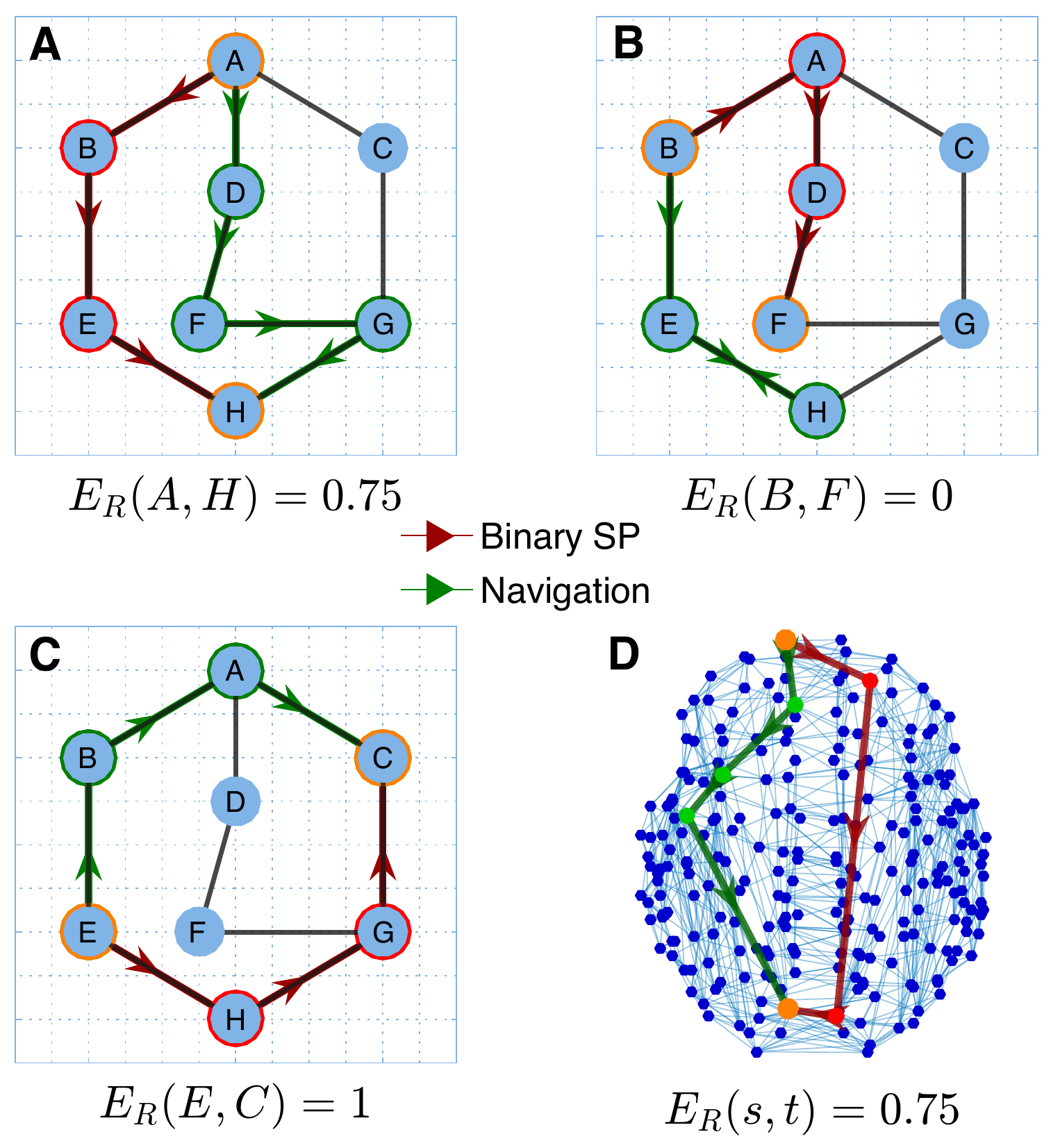}
	
	\caption{Illustrative examples of navigation (green) and shortest (red) paths from a source to target node (circled in orange) in a binary network. Grid indicates spatial embedding of the networks. Efficiency ratios ($E_{R}(i,j)$) are the ratio of the number of hops in the navigation path to the number of hops in the shortest path. \textbf{(A)} The shortest path between A and H has 3 hops (A-B-E-H) while navigation leads to a 4-hop path (A-D-F-G-H). Navigation routes information from A to H at 75\% of optimal efficiency. \textbf{(B)} Navigation fails to find a path from B to F, becoming trapped between E and H. \textbf{(C)} Both strategies lead to 3-hop paths, navigation routes information from G to B at 100\% of optimal efficiency. \textbf{(D)} Example of a successful navigation path in the human connectome that achieves 75\% efficiency.}
	
	\label{fig:toy}
	
\end{figure}

Navigation is a decentralized communication strategy that is particularly suited to spatially embedded networks \cite{boguna:2009, gulyas:2015}. Navigating a network involves following a simple rule: progress to the next directly connected node that is closest in distance to the target node, and stop if the target is reached (Fig. \ref{fig:toy}). To implement navigation, we defined the distance between pairs of nodes as the Euclidean distance between node centroids \cite{vertes:2012, betzel:2016}.  Importantly, navigation can fail to identify a path. This occurs when a navigation path becomes trapped between nodes without neighbors closer to the destination than themselves (Fig. \ref{fig:toy}B). The \textbf{success ratio} ($S_{R}$) measures the proportion of node pairs in a network that can be successfully reached via navigation.

Let $L \in \mathbb{R}^{N \times N}$ denote a matrix of connection lengths for a network comprising $N$ nodes, where $L_{ij}$ measures the length of the connection from node $i$ to $j$, and let $\Lambda$ denote the matrix of navigation path lengths. If node $i$ cannot navigate to node $j$, $\Lambda_{ij} = \infty$. Otherwise, $\Lambda_{ij} = L_{iu} + ... + L_{vj}$, where $\{u,...,v\}$ is the sequence of nodes visited during navigation. We define \textbf{navigation efficiency} as $E = 1 /(N^{2}-N) \sum_{i \neq j} 1 / \Lambda_{ij}$. Analogous to global efficiency \cite{latora:2001} ($E^{*} = 1 /(N^{2}-N) \sum_{i \neq j} 1 / \Lambda^{*}_{ij}$, where $\Lambda^{*}_{ij}$ is the shortest path length from node $i$ to $j$), both measures characterize the efficiency of information exchange in a parallel system in which all nodes are capable of concurrently exchanging information. In the same way as global efficiency can incorporate network disconnectedness, navigation efficiency incorporates unsuccessful navigation paths ($E_{ij} = 0$ if $i$ cannot reach $j$ under navigation). Therefore, $E$ quantifies both the number of failed paths and the efficiency of successful paths. The \textbf{efficiency ratio} 

\begin{equation} \label{eq:er}
E_{R} = \frac{1}{N^{2}-N} \sum_{i \neq j} \frac{\Lambda^{*}_{ij}}{\Lambda_{ij}}
\end{equation}

can be computed in order to compare navigation with shortest path routing. For any network, $E^{*} \geqslant E$ and thus $0 \leqslant E_{R} \leqslant 1$. The closer $E_{R}$ is to 1, the better navigation is at finding paths that are as efficient as shortest paths (Fig. \ref{fig:toy}).
We focus on binary ($E^{bin}_{R}$) and weighted ($E^{wei}_{R}$) navigation efficiency ratios, quantifying how efficient navigation paths are compared to shortest paths computed on binarized and weighted connectomes, respectively. In addition, we compute $E^{dis}_{R}$ to determine how close navigation paths are to routes that minimize the sum of physical (Euclidean) connection distances traversed between nodes.

\subsection*{Navigability of the human connectome.}

High-resolution diffusion MRI data from 75 healthy participants of the Human Connectome Project (HCP) \cite{van-essen:2013} were used to map structural brain networks at several spatial resolutions ($N=256,360,512,1024$). Whole-brain tractography was performed for each individual and the number of streamlines interconnecting each pair of nodes was enumerated to provide a measure of structural connectivity (SI). A group-level connectome was computed as the averaged of all individual connectivity matrices \cite{goni:2014}. Connection weights were remapped into binary, weighted and distance-based connection lengths to allow for the computation of communication path lengths (Materials and Methods).

\begin{figure*}
	\centering
	\includegraphics[width=1\linewidth,keepaspectratio]{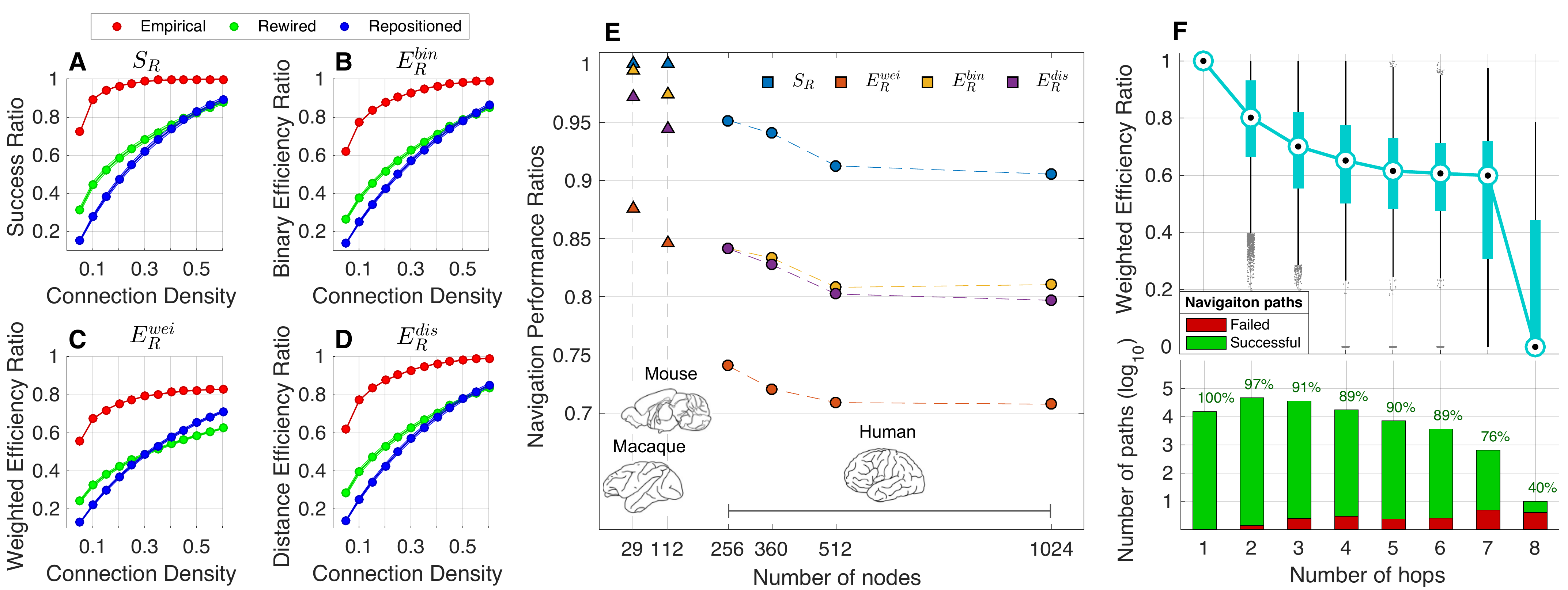}
	
	\caption{Navigability of mammalian connectomes. \textbf{(A-D)} Success ratio ($S_{R}$), binary efficiency ratio ($E^{bin}_{R}$), weighted efficiency ratio ($E^{wei}_{R}$) and distance efficiency ratio ($E^{dis}_{R}$) for human connectomes ($N=360$) at several connection density thresholds. Empirical measures (red) for group-averaged connectomes were compared to 1000 rewired (green) and spatially repositioned (blue) null networks. Shading indicates 95\% confidence intervals. \textbf{(E)} Same performance metrics shown across different parcellation resolutions of mammalian structural networks ($S_{R}$ shown in blue, $E^{wei}_{R}$ in orange, $E^{bin}_{R}$ in yellow and $E^{dis}_{R}$ in purple). Triangles denote non-human species while circles denote human data. Dashed lines denote the same connection density (15\%) across all human networks. \textbf{(F)} Navigability stratified by hop count ($N=360$ at 15\% connection density). Blue boxplots indicate the quartiles of $E^{wei}_{R}$ navigation paths benchmarked against shortest paths with matching hop count. Barplots show the number of shortest paths for a given hop count, with colors indicating the proportion of successful (green) and failed (red) navigation paths.}
	\label{fig:per}
\end{figure*}

Consistent with previous reports \cite{boguna:2009, gulyas:2015}, we found that navigation can successfully identify paths for the majority of nodes pairs comprising the human connectome ($S_{R}=$ 89\%, 94\%, 96\% for 10\%, 15\% and 20\% connection density, respectively, $N=360$; Fig. \ref{fig:per}A). Remarkably, navigation was only marginally less efficient than shortest paths (e.g., $E^{wei}_{R}=$ 72\%, $E^{bin}_{R}=$ 83\% and $E^{dis}_{R}=$ 83\%, for $N=360$ at 15\% connection density; Fig. \ref{fig:per}B-D), with navigation performance improving as connection density increased (note that navigation does not utilize connection weights, and thus $E_R^{wei}$ quantifies the extent to which navigation can blindly identify weighted shortest paths). Navigation remained efficient and successfully identified paths for the majority of node pairs across various parcellation resolutions (Fig. S3), with moderate decreases in success and efficiency ratios as the number of nodes increased ($S_{R}=$ 95\%, 91\%, 90\% and $E^{bin}_{R} =$ 84\%, 80\%, 81\% for $N=$ 256, 512, 1024, respectively, at 15\% connection density; Fig. \ref{fig:per}E). When stratified by hop-count, navigation performance remained high for long, multi-hop paths (61\% median $E^{wei}_{R}$ and 79\% median $E^{dis}_{R}$ benchmarked against 5-hop shortest paths; Fig. \ref{fig:per}F, S4), suggesting that navigation efficiency is not owing to directly connected node pairs, or node pairs that can be navigated in only a few hops.

Having established that the human connectome can be successfully and efficiently navigated, we next sought to determine whether efficient navigation is facilitated by the connectome's topology or its spatial embedding. Navigation performance was benchmarked against ensembles of random null networks in which either connectome topology was randomized \cite{maslov:2002} or the spatial position of nodes was shuffled. Navigability of both the topological and spatial null networks was markedly reduced in comparison to the empirical networks (46\% and 60\% decrease in $E^{bin}_R$; 48\% and 60\% decrease in $E^{wei}_R$, for topological and spatial randomization, respectively, at 15\% connection density; Fig. \ref{fig:per}A-D, blue and green curves), suggesting that navigation is jointly facilitated by both the connectome's topology and geometry. Simultaneous disruption of connectome topology and geometry yielded comparable results.

\subsection*{Navigability of non-human mammalian connectomes.}

Invasive tract-tracing studies provide high quality connectomes for a number of non-human species \cite{heuvel:2016}. We aimed to determine whether a 112-region mouse connectome (52\% connection density) \cite{rubinov:2015, oh:2014} and a 29-region macaque connectome (66\% connection density) \cite{markov:2014} were navigable. Navigation performed with $S_{R}=$ 100\% and near-optimal communication efficiency for both species ($E^{bin}_{R}=$ 99\%, 97\%, $E^{dis}_{R}=$ 97\%, 94\% and $E^{wei}_{R}=$ 87\%, 84\%, for the macaque and mouse, respectively; Fig. \ref{fig:per}E). As with the human connectome, navigation performance was significantly increased compared to the topologically rewired and spatial null networks  (all $P<10^{-4}$, with the exception of $P=0.012$ and $P=0.002$ for the macaque $S_{R}$ of topological and spatial null networks, respectively). 

The efficient navigation of connectomes across a variety of species, scales and mapping modalities suggests that the topology and spatially embedding of nervous systems is conducive to efficient decentralized communication, and randomization of either connectome topology or geometry is sufficient to render connectomes non-navigable.

\subsection*{Connectome topology maximizes navigation performance.}

Having found that randomizing connectome topology or geometry degraded navigation performance, we next sought to test whether navigation efficiency can be improved by progressively evolving the topology of the human connectome to either increase its regularity (clusterize) or increase its randomness and wiring cost. Randomization was performed by progressively swapping connections between randomly chosen node pairs, while preserving connection density and degree-distribution \cite{maslov:2002}. With sufficient iterations, this yielded the topologically randomized  null networks shown in Figs. \ref{fig:per}A-D (green curves). To clusterize topology, the same procedure was used with the additional constraint that each connection swap must lead to an increase in the overall clustering coefficient. Applying these procedures to evolve the connectome {\it in silico}, we generated two sets of networks that progressively tended towards different ends of an order spectrum, ranging from orderly and regular networks with a high clustering coefficient, to disordered and costly random networks (SI). Connectome geometry remained fixed.

\begin{figure}
	
	\centering
	\includegraphics[width=0.75\linewidth,keepaspectratio]{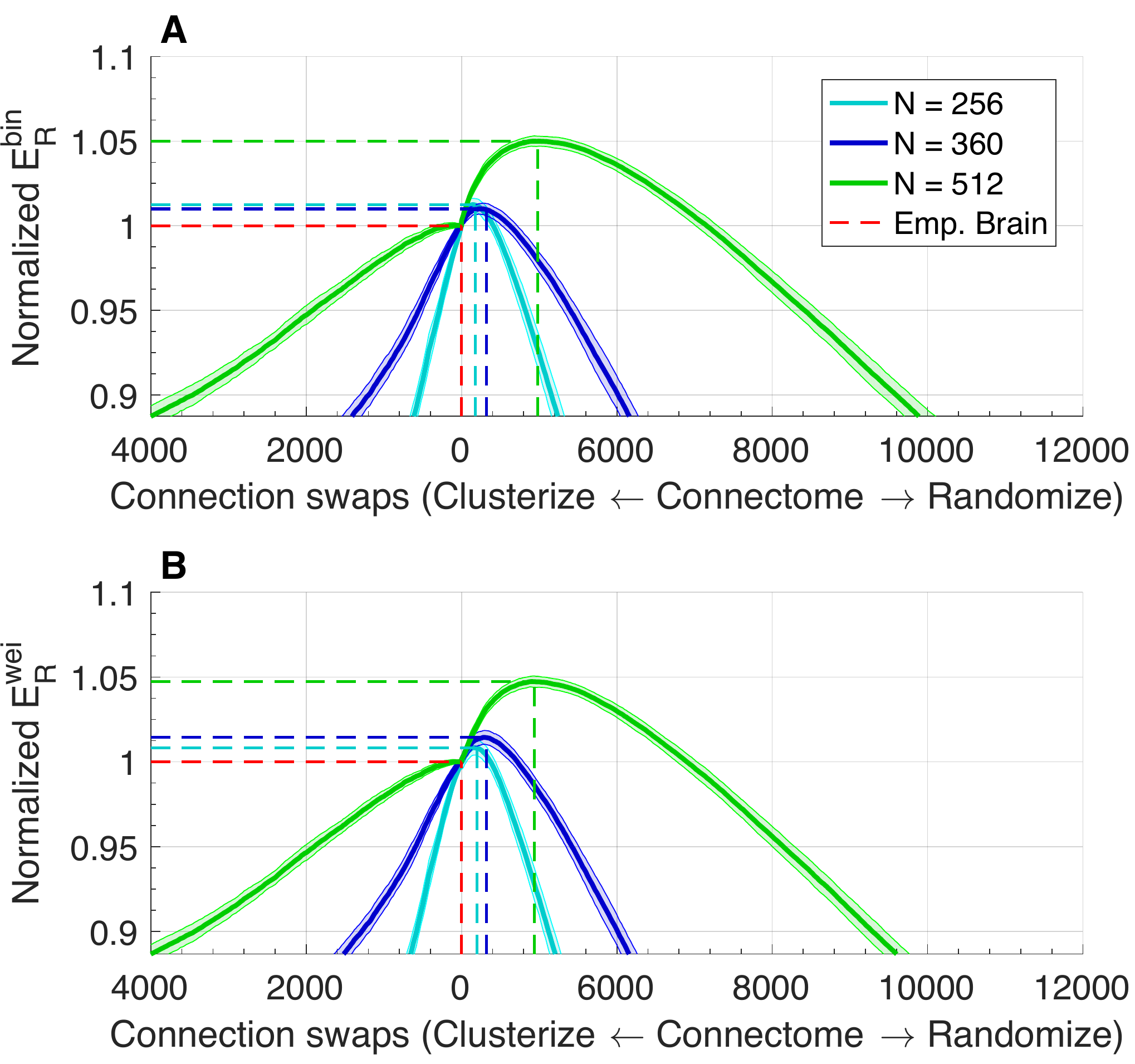}
	
	\caption{Navigation efficiency of clusterized and randomized networks, normalized by the human connectome's navigation efficiency. The horizontal axis shows the number of connection swaps performed in both randomizing (right of 0) and clusterizing rewirings (left of 0), with the human connectome located at 0 swaps. Curves indicate the mean values (inner line) and 95\% confidence intervals (outer shadow) of 100 runs of the randomization-clusterizing procedure for different parcellation resolutions at 15\% connection density. Dashed lines show performance peaks (vertical axis) and number of connection swaps (horizontal axis), with red indicating the values obtained for the empirical brain. \textbf{(A)} Normalized $E^{bin}_{R}$. \textbf{(B)} Normalized $E^{wei}_{R}$.}
	
	\label{fig:r2c}
	
\end{figure}

Clusterization was found to progressively decrease navigation efficiency, whereas slight randomization of connectome topology yielded networks with marginal increases in navigation performance (Fig. \ref{fig:r2c}). Specifically, peak $E^{bin}_{R}$ was, on average, 1.3\%, 1.1\% and 5.0\% more efficient than the connectome after 10.6\%, 9.4\% and 12.2\% of connections were randomly swapped, for $N=256,360,512$, respectively. Similarly, peak $E^{wei}_{R}$ was, on average, 0.8\%, 1.4\% and 4.7\% higher than empirical values, for $N=256,360,512$, respectively. Further randomization beyond these peaks resulted in deterioration of navigation efficiency. Therefore, the margin to improve navigation performance by evolving connectome topology is relatively small (<5\%), suggesting that brain networks are near-optimally poised between randomness and regularity to facilitate efficient, decentralized routing.

\subsection*{Navigation centrality.}

The number of shortest paths that traverse a node defines its betweenness centrality (BC), which is a measure that finds utility in identifying connectome hub nodes \cite{sporns:2007} and nodes mediating the bulk of neural communication \cite{harriger:2012}. We defined a new path-based centrality measure called navigation centrality (NC), which quantifies the number of successful navigation paths that traverse each node (SI). 

\begin{figure}
	
	\centering
	\includegraphics[width=0.95\linewidth,keepaspectratio]{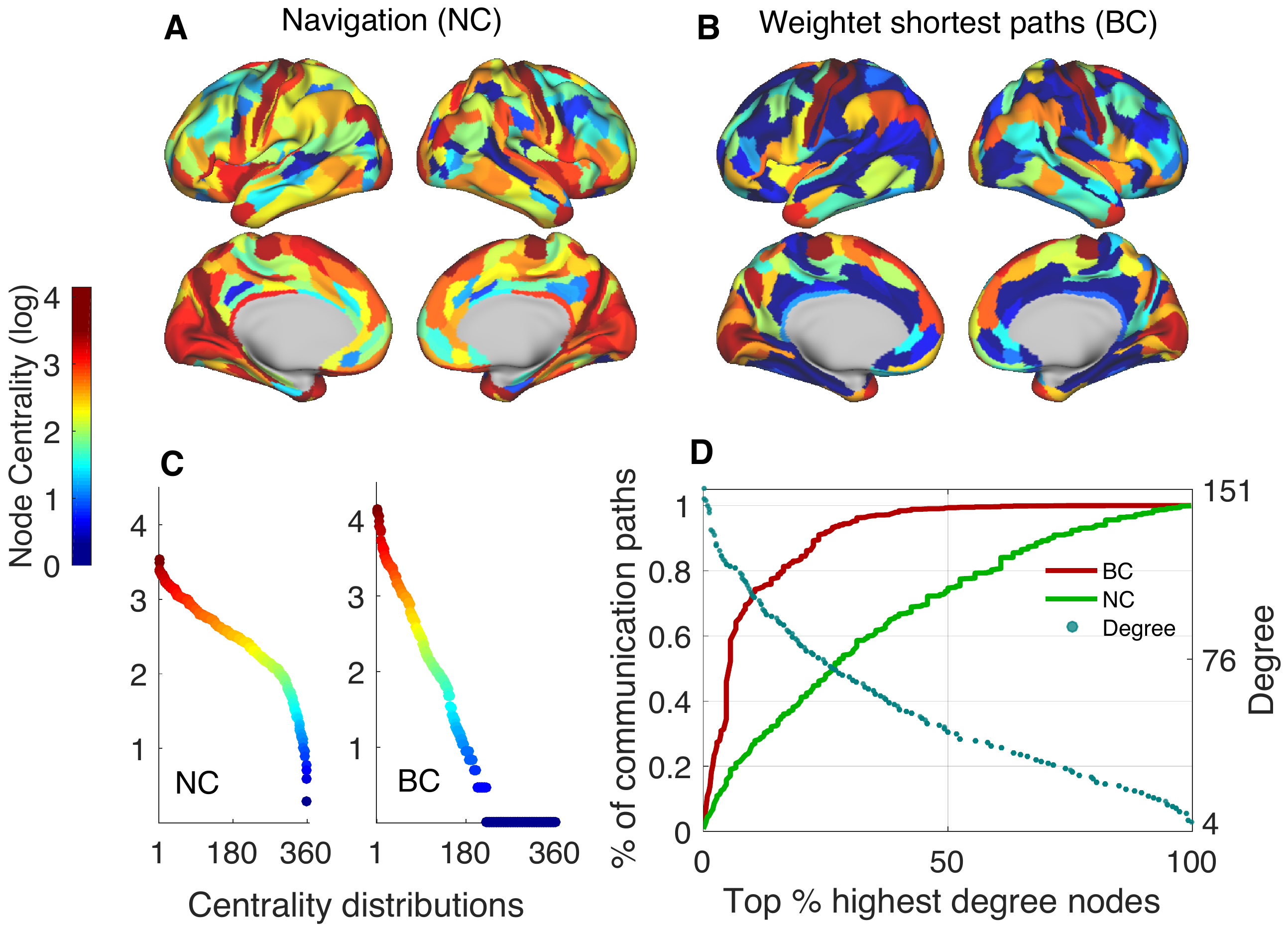}
	
	\caption{Comparison between navigation (NC) and weighted betweenness (BC) node centralities for $N=360$ at 15\% connection density. Centrality values are logarithmically scaled. NC \textbf{(A)} and BC \textbf{(B)} projected onto the cortical surface. \textbf{(C)} NC (right) and BC (left) sorted from highest to lowest values \textbf{(D)} Relationship between the cumulative sum of centrality measures and degree. The horizontal axis is a percentage ranking of nodes from highest to lowest degree (e.g., for $N=360$, the 10\% most connected nodes are the 36 nodes with highest degree). Solid curves (left-hand vertical axis) represent the cumulative sum of BC (red) and NC (green) over all nodes ordered from most to least connected, divided by the total number of communication paths in the network, indicating the fraction of communication paths mediated by nodes. Blue dots (right-hand vertical axis) show the original degree associated with each percentage of most connected nodes.}
	
	\label{fig:cen}
	
\end{figure}

We computed NC and BC for the human connectome (group average, $N=360$, at 15\% connection density), with BC based on weighted shortest paths. We found that both NC and BC spanned four orders of magnitude and were positively correlated (Pearson correlation coefficient $r=0.54$). Several regions were central to both shortest paths (BC) and navigation (NC), including portions of the left and right superior frontal gyrus, insula, central gyri and precuneus (Fig. \ref{fig:cen}A-B).  However, NC was more uniformly distributed across nodes compared to BC, suggesting that navigation utilizes network resources more homogeneously (Fig. \ref{fig:cen}C). High values of BC were found only in a small group of high-degree nodes ($r=0.86$ between logarithm of BC and degree), which mediated most of the network's communication routes. For instance, 99.3\% of all shortest paths travelled exclusively through the top 50\% most connected nodes (Fig. \ref{fig:cen}D). In contrast, although high-degree nodes showed high NC ($r=0.61$ between logarithm of NC and degree), medium- and low-degree regions were responsible for mediating a share of navigation paths, with the 50\% least connected nodes responsible for 26\% of navigation paths. Greater diversity in paths may lead to less communication bottlenecks and signal congestion \cite{misic:2014}, as well as stronger resilience against failure of network elements \cite{kaiser:2007, avena:2016}. Similar results were obtained for BC computed in binarized connectomes (Fig. S5). All correlation coefficients ($r$) were significant ($P<10^{-8}$).

\subsection*{Navigation and functional connectivity.}

Finally, we tested whether navigation path lengths can explain variation in functional connectivity across nodes pairs of the human connectome. The strength of functional connectivity between node pairs that are not directly connected can be attributed to signal propagation along multi-synaptic (multi-hop) paths \cite{honey:2009, goni:2014}. Therefore, if multi-hop neural communication is indeed facilitated by navigation, we hypothesized that navigation path lengths should be inversely correlated with functional connectivity strength. Resting-state functional MRI data from the same 75 participants of the HCP was used to map functional brain networks. A group-averaged functional network was obtained by averaging FC across all participants (SI).

Given the geometric nature of navigation, an intuitive definition of navigation path length is the total distance travelled from one node to another along the navigation path (SI). Navigation path lengths and FC were significantly associated across node pairs, with Pearson correlation coefficient $r=-0.32$ and $r=-0.43$, for the whole brain and right hemisphere, respectively, for $N=360$ at 15\% connection density (Fig. S6B, S6D). This relationship was consistent across density thresholds, with navigation path lengths yielding stronger correlations with FC than shortest path lengths ($r=-0.35$, right hemisphere) and the direct Euclidean distance between node pairs ($r=-0.40$, right hemisphere; Fig. S6A, S6C). Navigation path lengths remained a significant predictor of FC when including Euclidean distance as a covariate. All correlation coefficients ($r$) were significant ($P<10^{-8}$).

\section*{Discussion}

This study investigated navigation as a model for large-scale neural communication. Using a novel measure of navigation efficiency, we evaluated the navigability of a range of mammalian connectomes that were analyzed as binary, weighted and distance-based networks. We found multiple lines of evidence suggesting that the topology (wiring) and spatial embedding (geometry) of nervous systems is conducive to efficient navigation. The margin to evolve connectomes to improve navigation performance was minimal. Our new measure of navigation centrality indicated that navigation utilizes network resources more uniformly compared to shortest paths. Moreover, variation in functional connectivity across pairs of nodes was well explained by navigation path lengths. We conclude that navigation is a viable neural communication strategy that does not mandate the biologically unrealistic assumptions inherent to shortest paths, nevertheless achieving near-optimal routing efficiency.

Navigation performance was assessed for binary, weighted and distance-based connectomes, each emphasizing a distinct attribute of network communication. While binary and weighted brain networks are commonly used in connectomics \cite{fornito:2016}, distance-based connectomes were introduced to provide a geometric benchmark for navigation efficiency. Interestingly, navigation paths were simultaneously efficient in all three regimes, suggesting that neural signaling may favor communication routes that combine few synaptic crossings (binary), high axonal strength and reliability (weighted) and short propagation distances (distance-based).

Navigability increased with connection density. Trivially, if all node pairs are connected, both navigation and shortest paths consist of direct connections. While dense human and non-human connectomes were near-optimally navigable (97\%--100\% $E^{bin}_{R}$ at 50\%--60\% connection density), density-matched null networks showed significantly reduced navigability. In addition, progressive rewiring of connectome topology revealed that the human brain resides close to peak navigation performance in a topological order (regularity-randomness) spectrum (improvement margin $\sim$1\% for $N=360$), suggesting that brain organization is shaped to facilitate efficient decentralized communication. While our focus on the order spectrum was motivated by canonical work on brain and real-world networks \cite{watts:1998, bullmore:2012}, other dimensions of connectome attributes, such as modular--centralized, could be investigated \cite{betzel:2016, jarman:2017}.

Navigation path lengths were defined in terms of the Euclidean distance travelled along navigation paths. The significant association between navigation and FC is further evidence that neural communication is not necessarily constrained to optimal routes \cite{goni:2014}. Navigation path lengths remained a significant predictor of FC when including Euclidean distance as a covariate. Thus, the association between navigation and FC cannot be entirely attributed to navigation path lengths approximating the physical distance between nodes, indicating that the interplay between topological and geometric distances may contribute to the relationship between brain structure and function.

\subsection*{Connectome geometry, topology and communication.}

Several recent studies have drawn attention to the link between geometry and topology in neural and other real-world networks \cite{boguna:2009, vertes:2012, betzel:2016}. Our findings indicate that the interplay between connectome geometry and topology may play an important role in facilitating efficient neural communication. The association between network geometry and topology \cite{betzel:2016, roberts:2016} contributes to the appearance of topological attributes conducive to navigation \cite{boguna:2009, allard:2017}; while the spatial positioning of nodes guides navigation of the topology, facilitating efficient decentralized communication.

Successful navigation has been linked to small-world topologies that combine spatially separated hubs and high clustering coefficients \cite{kleinberg:2000, boguna:2009}. Hub-to-hub long-range connections facilitate rapid information transfer across distant regions of the brain (although see \cite{betzel:2017} for a counterpoint), while high clustering enables navigation to home in on specific destinations \cite{boguna:2009, heuvel:2012} (see Fig. S7). This mechanism for information transfer is consistent with observations suggesting that nervous systems are small world networks, balancing integration supported by long-range connections and segregation due to locally clustered modules \cite{bullmore:2012}.

\subsection*{Biological plausibility and communication efficiency.}

To the date, the study of brain network organization has been anchored to the assumption of communication under shortest path routing \cite{rubinov:2010, fornito:2016}. Two examples are the characterization of the brain as a small-world network \cite{watts:1998, bassett:2006} and the use of global efficiency as a measure of network integration \cite{bullmore:2009, bullmore:2012}. These topological properties are derived from the shortest paths between all pairs of nodes. We found that navigation can approximate the overall efficiency of shortest paths without requiring centralized knowledge of global network topology. Thus, we provide reassurance that previous findings obtained under the shortest paths assumption remain pertinent, despite the biologically implausible requirements for the computation of optimal routes; and reaffirm the importance of the brain's small-world architecture in the light of decentralized routing schemes.

Alternative decentralized communication models such as diffusion processes \cite{goni:2013, goni:2014}, communicability \cite{estrada:2008} and spreading dynamics \cite{misic:2015} have been proposed as biologically plausible alternatives, and merit further study. The metabolic cost of these alternative models is likely to be higher than navigation because they assume that neural signaling is broadcast over the entire network, as opposed to being routed to a destination via a single path \cite{bullmore:2012, avena:2017}. Moreover, recent evidence suggests that connectome topology is not optimized for efficient communication via diffusion \cite{avena:2014}.

\subsection*{Future directions.}

Navigation depends on the assumption that network nodes possess information about the relative spatial positioning between their direct neighbours and a target node. Further work is needed to establish whether this assumption is biologically plausible. Nevertheless, this assumption is more realistic than the global knowledge of network topology required by shortest paths routing.

Non-invasive neuroimaging modalities generally do not enable explicit observation of information transfer between distant neural elements. In the future, brain stimulation techniques might be used to evaluate evidence for competing communication strategies by means of electrophysiological tracking of local perturbations \cite{gollo:2017}. Alternatively, virtual lesioning of network elements along navigation or shortest paths could be applied to investigate potential changes in simulated patterns of functional co-activation between nodes \cite{alstott:2009, grayson:2016}.

\section*{Materials and Methods}
Details on human, macaque and mouse network datasets, including human diffusion and functional MRI data acquisition and preprocessing pipelines, are described in {\it SI Materials and Methods, Connectivity data}. {\it SI Materials and Methods, Network analysis} provides definitions, formulae and details on network modelling and analyses. Further supporting and replication analyses are presented in {\it SI Supplementary analyses}.

\subsection*{Connection weights and lengths.}
A weighted connectome can be expressed as a matrix $W \in \mathbb{R}^{N \times N} $, where $W_{ij}$ is the connection weight between nodes $i$ and $j$. Connection weights are a measure of similarity or affinity, denoting the strength of the relationship between two nodes (e.g., streamline counts in tractography or fraction of labelled neurons in tract tracing). Typically, the matrix of connection lengths $L$ is computed from $W$ by means of a weight-to-length transformation that monotonically remaps strong weights into short lengths and weak weights into long lengths. This way, connection lengths denote the dissimilarity, distance or travel cost between nodes. Defining nodal relationships in terms of lengths allows for the computation of short communication paths.

Connection length matrices were computed as follows. In the binary case, $L^{bin}\in[0,1]$ denoting only the presence or absence of a connection in $W$. In the weighted case, $L^{wei}=-\log_{10}(W/max(W))$ ensuring a monotonic weight-to-length transformation that produces log-normally distributed  connection lengths and attenuates extreme connection weights \cite{goni:2014, avena:2016} (for human connectomes weighted by streamline counts, $L^{wei}=-\log_{10}(W/max(W)+1)$ to avoid remapping the largest weight to a length of zero.) In the distance-based case, connections lengths denote the physical distance between directly connected nodes, with $L^{dis}=D \odot L^{bin}$, where D is the Euclidean distance matrix between node pairs. Hence, shortest paths computed on $L^{bin}$, $L^{wei}$ and $L^{dis}$ minimize the number of traversed connections (hops), maximize the strength and reliability of traversed connections and minimize the physical distance traversed between nodes, respectively.

\subsection*{Navigation implementation.}
For a network with $N$ nodes, navigation routing from node $i$ to $j$ was implemented as follows. Determine which of $i$'s neighbors is closest (shortest Euclidean distance) to $j$ and progress to it. Repeat this process for each new node until: i) $j$ is reached---constituting a successful navigation path---or ii) a node is revisited---constituting a failed navigation path.

Navigation paths are identified based on network topology and the spatial positioning of nodes, and thus independent from how connection lengths are defined. Navigation paths lengths, however, are the sum of connection lengths comprised in navigation paths, and will vary depending on the definition of $L$. The matrix of navigation path lengths $\Lambda$ was computed by navigating every node pair. Note that $\Lambda$ is asymmetric, requiring $N^{2}-N$ navigation path computations. Using different connection length measures ($L^{bin}, L^{wei}, L^{dis}$), we computed binary ($\Lambda^{bin}$), weighted ($\Lambda^{wei}$) and distance-based ($\Lambda^{dis}$) navigation path lengths. Navigation efficiency ratios ($E_{R}^{bin}$, $E_{R}^{wei}$, $E_{R}^{dis}$) were computed by comparing navigation path lengths to shortest path lengths ($\Lambda^{bin^{*}}$, $\Lambda^{wei^{*}}$, $\Lambda^{dis^{*}}$) using Eq. \ref{eq:er}.

\subsection*{Data sharing.} The analyzed human, macaque and mouse datasets are publicly available. Our implementation of navigation routing is available at \url{https://github.com/caioseguin/connectomics/}.

\bibliography{nav_bib_arxiv}
\bibliographystyle{ieeetr}
	
\end{document}


\maketitle

\section*{Overview}

This supporting document provides details on the acquisition and preprocessing of the analyzed connectivity data. We also provide descriptions of the network measures and methods applied throughout this work. Finally, we include supplementary and replication analyses that indicate the robustness and universality of the results reported in the main manuscript.

\section*{Materials and Methods}

\subsection*{Connectivity data}

\subsubsection*{Mouse.}

The Allen Institute for Brain Science mapped the mesoscale topology of the mouse nervous system by means of anterograde axonal injections of a viral tracer \cite{oh:2014}. Using two-photon tomography, they identified axonal projections from the 469 injections sites to 295 target regions. Building on these efforts, Rubinov and colleagues constructed a bilaterally symmetric whole-brain network for the mouse, comprising $N=112$ cortical and subcortical regions with 53\% connection density \cite{rubinov:2015}. Connections represent interregional axonal projections and their weights were determined as the proportion of tracer density found in target and injected regions. Connection weights followed a log-normal distribution and graph-theoretical analyses of the constructed topology revealed many organisational similarities to the human connectome.

\subsubsection*{Macaque.}

Markov and colleagues applied 1615 retrograde tracer injections to 29 of the 91 areas of the macaque cerebral cortex, spanning occipital, temporal, parietal, frontal, prefrontal and limbic regions \cite{markov:2013, markov:2014}. This resulted in a $29 \times 29$ directed interregional sub-network of the macaque coritco-cortical connections. Connection weights were estimated based on the number of neurons labelled by the tracer found in source and target regions, relative to the amount found in whole brain. In line with other tract-tracing studies, their network showed high connection density (66\%) and weights were distributed log-normally.

\subsubsection*{Human data acquisition and preprocessing.}

Minimally preprocessed diffusion weighted MRI data from 75 unrelated healthy participants (age 22--35, 40 females) was obtained from the Human Connectome Project (HCP) \footnote{\tiny HCP participants ID: 100206, 100307, 100408, 101006, 101107, 101309, 101915, 102109, 102513, 102614, 102715, 102816, 103111, 103212, 103414, 103515, 103818, 104012, 105014, 105115, 105216, 105620, 106016, 106319, 106521, 106824, 107018, 107321, 107422, 107725, 108020, 108121, 108222, 108323, 108525, 108828, 109123, 109830, 110007, 110411, 110613, 111009, 111211, 111312, 111413, 111716, 112112, 112314, 112516, 112920, 113215, 113316, 113619, 113922, 114217, 100610, 102311, 104416, 118225, 105923, 111514, 114823, 115017, 125525, 128935, 131722, 140117, 146129, 146432, 155938, 156334, 157336, 158035, 158136, 162935.} \cite{van-essen:2013}. Spin-echo planar diffusion weighted imaging was performed in a customized Siemens Skyra 3T scanner according to the following parameters: 5520 ms repetition time, 89.5 ms echo time, 78 degree flip angle, 160 degree refocusing flip angle, 210 $\times$ 180 mm filed of view, 168 $\times$ 144 matrix, 1.25 mm isotropic voxels, 3 shells of b=1000, 2000 and 3000 s/mm2 and approximately 90 diffusion weighted directions per gradient table. The obtained diffusion images were submitted to the HCP diffusion preprocessing pipeline \cite{glasser:2013} consisting of: 1) $b_{0}$ intensity normalization, 2) top-up EPI distortion correction, 3) Eddy current distortions and subject motion correction, 4) gradient distortion correction and 5) resampling to 1.25mm native structural space. Refer to \cite{glasser:2013, sotiropoulos:2013} for further details on HCP diffusion MRI acquisition and preprocessing.

Minimally preprocessed resting-state functional MRI data from the same 75 participants was also obtained from the HCP. Data acquisition consisted of four 14m33s runs, two runs (right-to-left and left-to-right phase encodings) in one session and two in another session, with eyes open with relaxed fixation on a projected bright cross-hair on a dark background (and presented in a darkened room). Data collection was performed in a customized Siemens Skyra 3T scanner according to the following parameters: gradient-echo EPI sequence, 720 ms TR, 33.1 ms TE, 52 degree flip angle, 208 $\times$ 180 mm FOV, 104 $\times$ 90 matrix, 2.0 mm slice thickness, 72 slices, 2.0 mm isotropic voxels, 8 multiband factor and 0.58 ms echo spacing. Acquired images were preprocessing according to the HCP functional preprocessing pipeline \cite{glasser:2013}, which involves: 1) spatial and gradient distortion corrections, 2) correction of head movement, 3) intensity normalization 4) single spline re-sampling of EPI frames into 2mm isotropic MNI space and 5) HCP's FIX+ICA pipeline for the removal of temporal artefacts. Refer to \cite{glasser:2013,smith:2013} for further details on HCP resting-state functional MRI acquisition and preprocessing.

\subsubsection*{Human cortical parcellation.}

Connectome analyses are sensitive to the number of nodes used to reconstruct brain networks \cite{zalesky:2010}. In order to assess the validity of our results across connectomes defined over different granularities of cortical segmentations, we generated parcellations of the human cerebral cortex containing $N=256,512,1024$ regions. In addition, we also mapped connectomes using the Glasser Atlas \cite{glasser:2016}, a multi-modal cortical parcellation built from the combination of structural, diffusion and functional imaging data from HCP participants.

We developed an algorithm to generate high-resolution cortical parcellations by means of sub-dividing an input low-resolution parcellation. Starting from a well established anatomical atlas ($N=68$) \cite{desikan:2006} defined on FreeSurfer's (\url{https://surfer.nmr.mgh.harvard.edu/}) {\it fsaverage} cortical sheet, each surface region was subdivided according to three criteria: (i) the resulting number of regions is specified by the user as a power of 2; (ii) subdivisions must respect the original anatomical boundaries, so that all resulting parcels can be attributed to a single anatomical structure; and (iii) each resulting sub-region has approximately the same surface area. Criterion (i) allows for a systematic investigation of the robustness of graph theory measures to variations in parcellation resolution. Criterion (ii) ensures that the resulting parcellation comprises biologically meaningful regions that can be assigned to a single anatomical input region. Finally, criterion (iii) minimizes tractography biases related to the surface area and volume of regions. Importantly, defining the parcellations over the cortical surface ensures that every grey matter region has an interface with white matter, making it accessible to tractography streamlines.

\begin{figure}
	\centering
	\subfloat[$N=68$]{
		\includegraphics[scale=0.35]{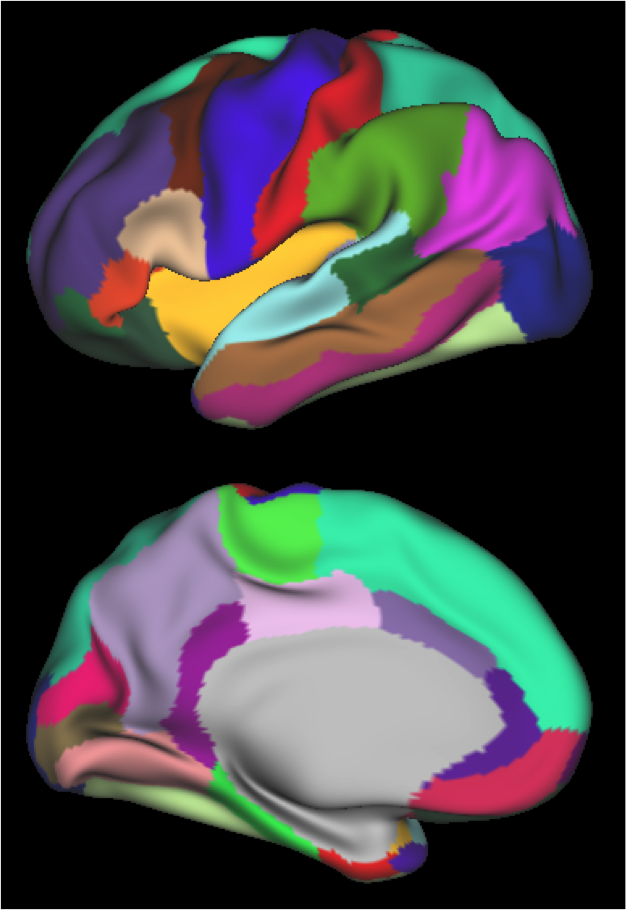}
		\label{fig:par_dk}
	}\hspace{1.5em}
	\subfloat[$N=512$]{
		\includegraphics[scale=0.35]{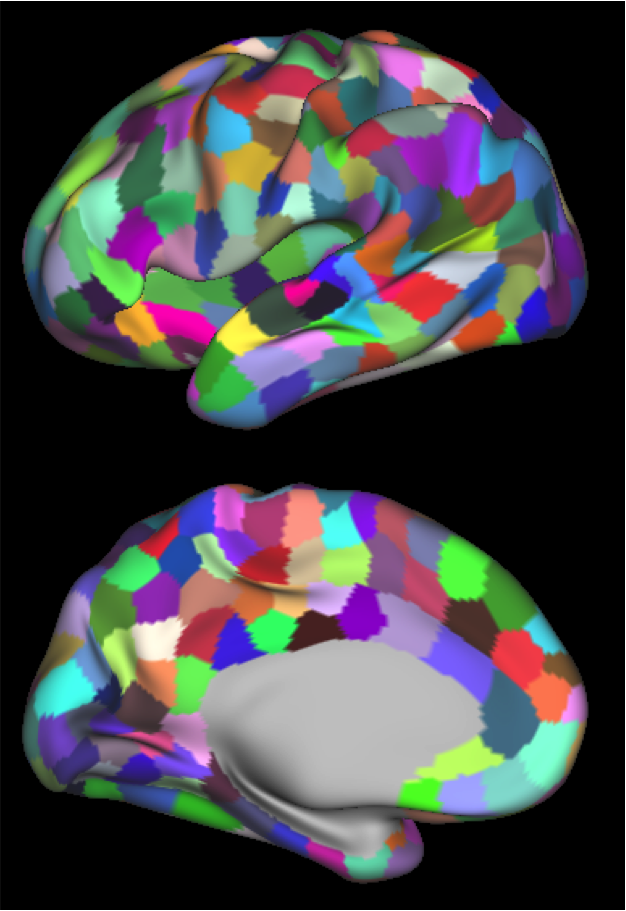}
		\label{fig:par_512}
	}\hspace{1.5em}
	\subfloat[$N=1024$]{
		\includegraphics[scale=0.35]{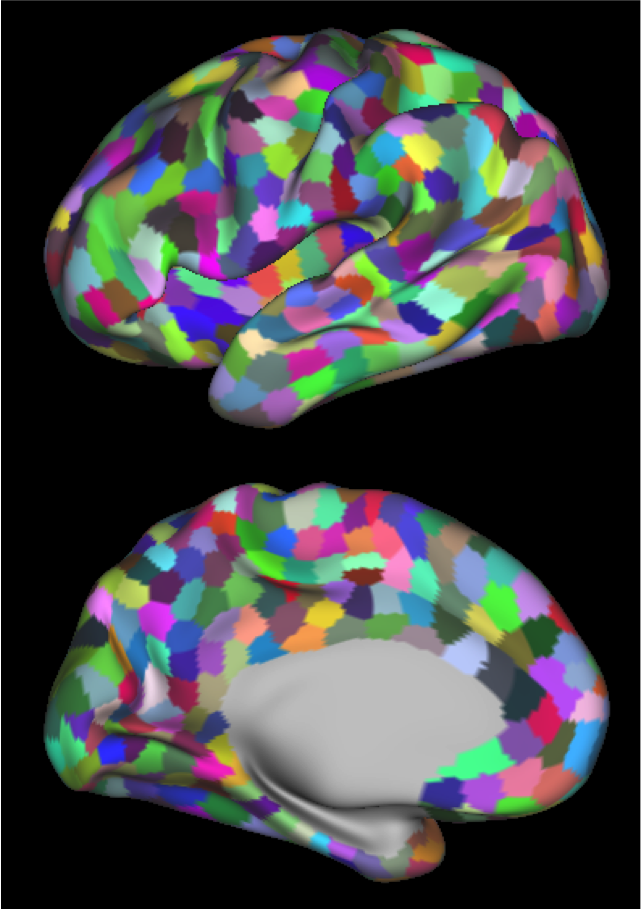}
		\label{fig:par_1024}
	}
	\caption{Cortical surface parcellation (left hemisphere). \textbf{(A)} Input low-resolution parcellation delineating 68 cortical regions. Output high-resolution parcellations with 512 \textbf{(B)} and 1024 \textbf{(C)} regions. All output sub-regions are contiguous, tile-shaped and have approximately the same surface area, while respecting the anatomical boundaries defined by the original 68 region atlas.}
	
	\label{fig:par}
	
\end{figure}

Input low-resolution regions were divided based on their size (number of vertices in the cortical sheet mesh) and the desired number of high-resolution output regions. Large input regions originated many output regions, while small input regions gave rise to few output regions. K-means++ \cite{arthur:2007} applied to the spatial coordinates of vertices was used to perform region division, leading to evenly sized, spatially contiguous, tile-shaped output regions (Fig. \ref{fig:par}). The resulting cortical surface parcellations were registered to native subject space and projected into parcellation volumes using Connectome Workbench (\url{https://www.humanconnectome.org/software/connectome-workbench}, \cite{marcus:2011}).

While cortical regions are commonly defined in terms of a surface mesh, representations of subcortical structures remain voxel-based \cite{glasser:2013}. Parcellating subcortical regions is therefore more challenging and requires different methodologies to those that we have applied to the cortex. For this reason, subcortical regions were not included in the connectomes reconstructed in this study.

\subsubsection*{Human network mapping.}

Structural networks (connectomes) were mapped for each individual using a deterministic tractography \cite{maier-hein:2017} pipeline from the MRtrix3 software (\url{http://www.mrtrix.org/}, \cite{tournier:2012}). Diffusion tensors were estimated from the preprocessed diffusion weighted images using a iteratively reweighted linear least squares estimator. White matter masks was extracted from structural segmentation files provided by the HCP. White matter boundaries were dilated by 1 voxel to fill potential gaps between grey and white matter caused by slight imperfections in subject registration. Streamlines were uniformly seeded from the white matter mask and tracked along diffusion tensor directions (eigenvectors) until they exited the mask into grey matter or reached a voxel with low fractional anisotropy (FA) (FACT tracking algorithm, $5 \times 10^{6}$ streamlines, 0.5 mm tracking step-size, 400 mm maximum streamline length and 0.1 FA cutoff for termination of tracks). Connection strength between any pair of regions was determined as the number of streamlines with extremities located in the regions, resulting in $N \times N$ weighted connectivity matrices. In order to capture general patterns of brain organization and filter out idiosyncratic variation, a group average connectome was computed by averaging the individual connectomes of all 75 participants. Subsequent network analyses were carried out on the group-averaged connectome.

Functional networks were mapped using command line tools from Connectome Workbench. For each subject, four runs of preprocessed resting-state fMRI volumes were averaged together and projected onto the subject's cortical surface. The functional activation profile of each region was described as a single time series by averaging the signal of all the vertices comprised in the region. Pairwise functional connectivity was computed by means of the Pearson correlation coefficient between regional time series, resulting in $N \times N$ weighted connectivity matrices. A group-averaged functional network comprising the 75 individual connectivity matrices was computed and used in subsequent analyses.

\subsection*{Network analysis}

\subsubsection*{Connectome thresholding.} Density-based thresholding of connection weights was applied to the group-averaged connectome in order to filter out potentially noisy and spurious connections \cite{fornito:2013}. A density threshold of 15\% consists of keeping the top 15\% highest weighted connections and deleting the remaining ones. Since the true connection density of the brain remains unknown \cite{fornito:2016}, we sought to replicate our key analyses for a range of 5\% to 60\% connection density thresholds, at 5\% intervals.

\subsubsection*{Null network models.}

Four null network models were used in this work: (M1) topologically randomized (rewired) networks, (M2) spatially randomized (repositioned) networks, (M3) progressively randomized networks and (M4) progressively clusterized  networks. Null models (M1) and (M2) were employed in the sections {\it Navigability of the human connectome} and {\it Navigability of non-human mammalian connectomes}, while (M3) and (M4) were used in section {\it Connectome topology maximizes navigation performance}.

Topological randomized networks (M1) were computed by rewiring the connectome using the Brain Connectivity Toolbox (\url{https://sites.google.com/site/bctnet/}, \cite{rubinov:2010}) implementation of the Maslov-Sneppen algorithm \cite{maslov:2002}. Rewiring was carried out by swapping each connection once (on average) while preserving the network original degree distribution and connectedness. Spatially randomized networks (M2) had the spatial coordinates or their nodes randomly swapped with each other, while maintaining the topology unaltered.

For reasons of computational tractability, navigation performance in the human connectome ({\it Navigability of the human connectome}) was benchmarked against ensembles of 1000, 1000, 500 and 100 topologically randomized networks (M1), per connection density threshold, for $N=256,360,512,1024$, respectively. Ensembles of the same size were generated for spatially randomized networks (M2). The benchmarking of non-human connectomes ({\it Navigability of non-human mammalian connectomes}) was done by constructing ensembles of $10^4$ topologically randomized networks (M1) and $10^4$ spatially randomized networks (M2). Non-parametric P-values for $S_{R}$, $E^{bin}_{R}$, $E^{wei}_{R}$ and $E^{dis}_{R}$ were computed as the proportion of times null networks outperformed the empirical connectomes.

Progressively randomized networks (M3) were obtained by the same rewiring process as topologically randomized networks (M1), consisting simply of partially rewired networks for which only a fraction of connections have been swapped. Progressively clusterized networks (M4) were produced by an adaptation of the (M3) procedure that only considers connection swaps that lead to an increase in the overall clustering coefficient of the network. The randomizing--clusterizing routine ({\it Connectome topology maximizes navigation performance}) was repeated 100 times for $N=256,360,512$, with $E^{bin}_{R}$ and $E^{wei}_{R}$ re-computed once every 5 connection swaps.

\subsubsection*{Navigation centrality.}
	
Binary and weighted betweenness centralities were computed using the Brain Connectivity Toolbox, representing the absolute number of shortest paths that pass through each node of the network. Similarly, the navigation centrality of node $i$ defined as $c(i) = \sum_{s \neq i \neq t} \pi_{st}(i)$, where $\pi_{st}=1$ if the successful navigation path from $s$ to $t$ passes through $i$ and $\pi_{st}=0$ otherwise.

\subsubsection*{Navigation and functional connectivity.}

Our implementation of navigation identifies paths based on the Euclidean distance between network nodes. Thus, a natural definition of navigation path length is to consider the physical distance traversed from one node to another. For our FC analysis, we defined navigation path length from $i$ to $j$ as  $\Lambda^{nav}_{ij}=log_{10}(D_{iu} + ... + D_{vj})$, where $\{u,...,v\}$ is the sequence of nodes visited during navigation from $i$ to $j$ and $D$ is the Euclidean distance matrix between node pairs.

\begin{figure}[ht]	
	\centering
	\includegraphics[scale = 0.45]{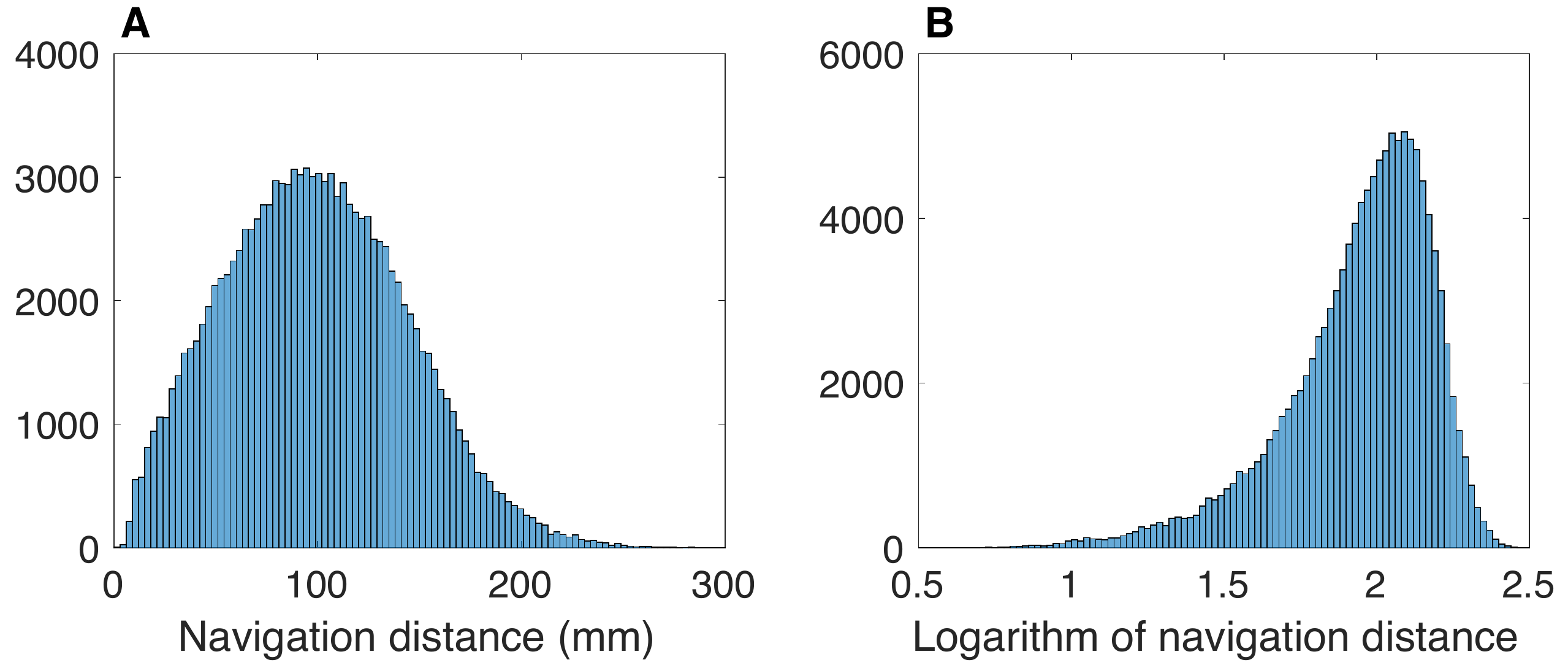}
	
	\caption{Distribution of raw \textbf{(A)} and log-transformed \textbf{(B)} navigation path lengths between all pairs of nodes ($N=360$ at 15\% connection density threshold).}
	
	\label{fig:si_ndd}
	
\end{figure}

The logarithmic transformation leads to a better differentiation of the distance covered by short navigation paths (Fig. \ref{fig:si_ndd}), leading to an expansion of the left tail (short navigation distances) of the navigation path length distribution and a compression of the right tail (long navigation distances). The transformation leads to stronger correlations with FC, suggesting that a nuanced characterization of short communication paths might play an important role in understanding the relationship between structural and functional brain networks.

\section*{Supplementary analyses}

\subsubsection*{Distance scaling of connection weights.} A common practice in connectomics studies based on white matter tractography is to scale streamline counts by average fiber lengths \cite{hagmann:2008}. This is motivated by the notion that certain tractography algorithms are biased towards overestimating the streamline count of long fiber bundles. We chose not to perform distance-based scaling, since this would impart information on the distance between nodes in each network connection weight, potentially leading to an artificial improvement in navigation performance. However, it is then possible that the observed values of $E^{wei}_R$ are a result of the uncorrected tractography bias towards long fiber bundles, since long range connections would be more efficient for both streamline count and distance-based path length regimes. In order to test for this issue, we recomputed navigation performance after dividing streamline counts by the Euclidean distance between nodes. As initially hypothesized, this transformation lead to improved navigation performance, with $S_{R}=96\%$, $E^{bin}_R=86\%$, $E^{wei}_R=81\%$ and $E^{dis}_R=84\%$, for $N=360$ at 15\% connection density.
	
\begin{figure}
		\centering
		\subfloat[$N=256$]{
			\includegraphics[scale=0.35]{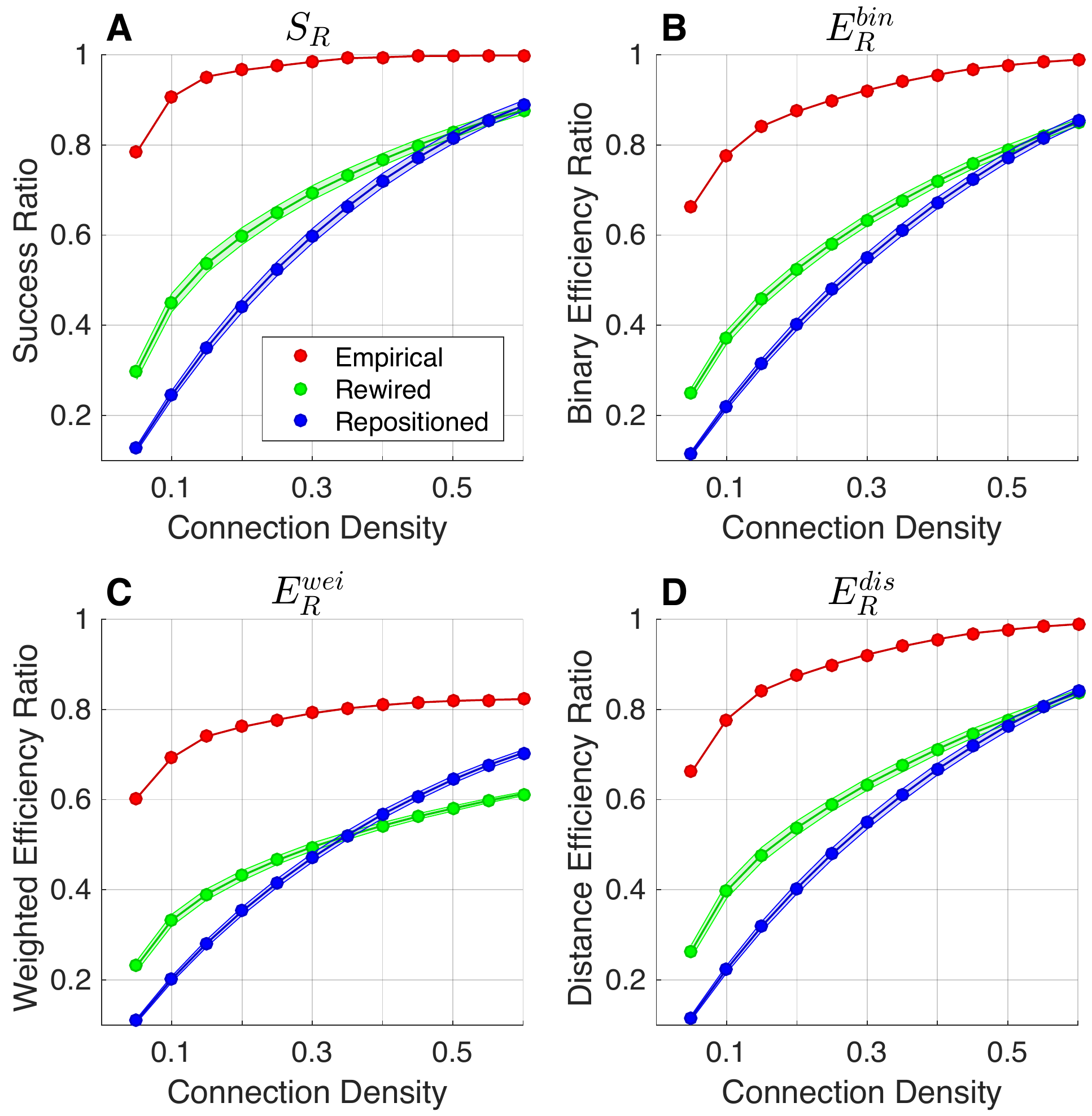}
			\label{fig:si_per_256}
		}\hspace{1.5em}
		\subfloat[$N=512$]{
			\includegraphics[scale=0.35]{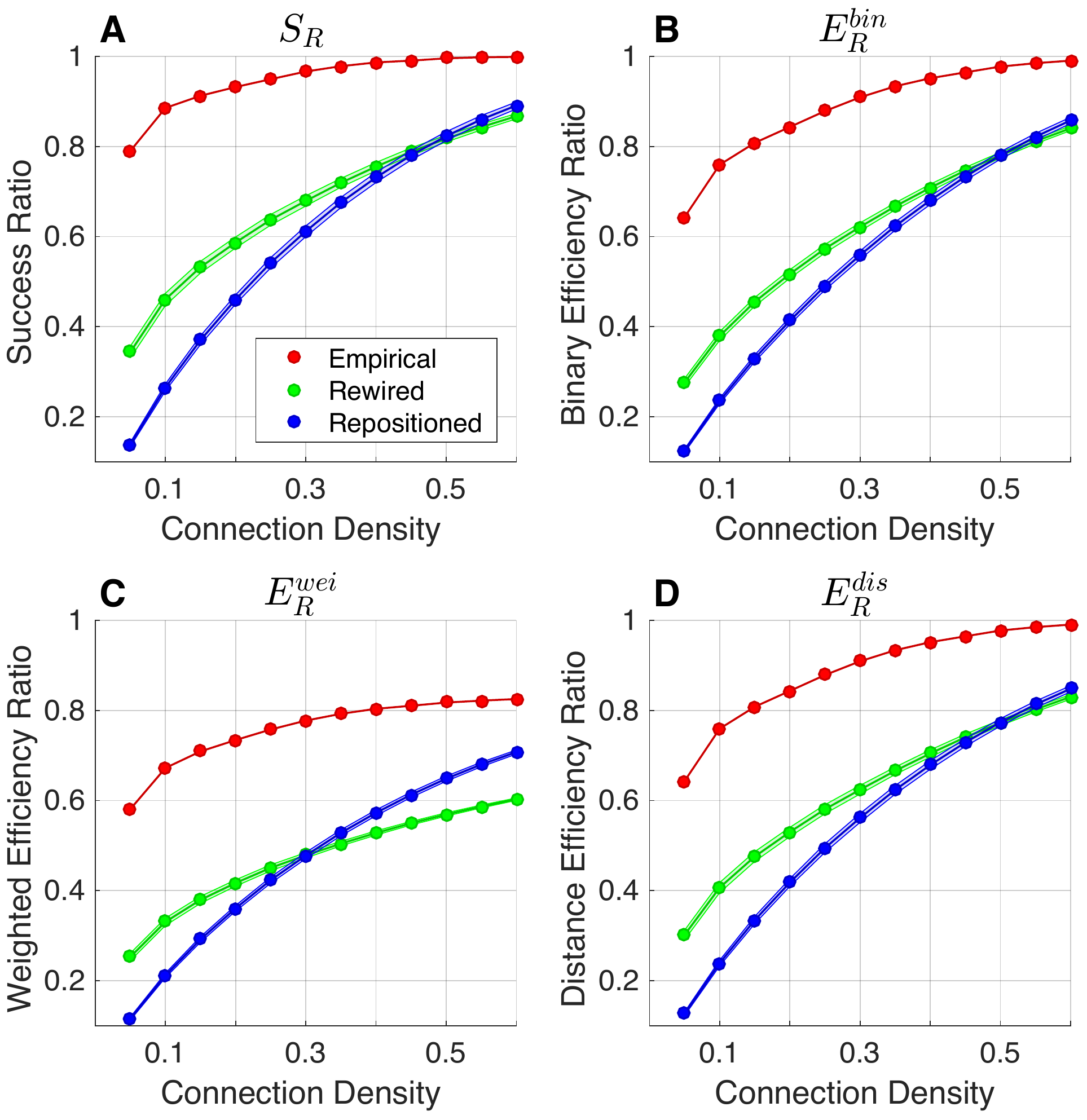}
			\label{fig:si_per_512}
		}\hspace{1.5em}
		\subfloat[$N=1024$]{
			\includegraphics[scale=0.35]{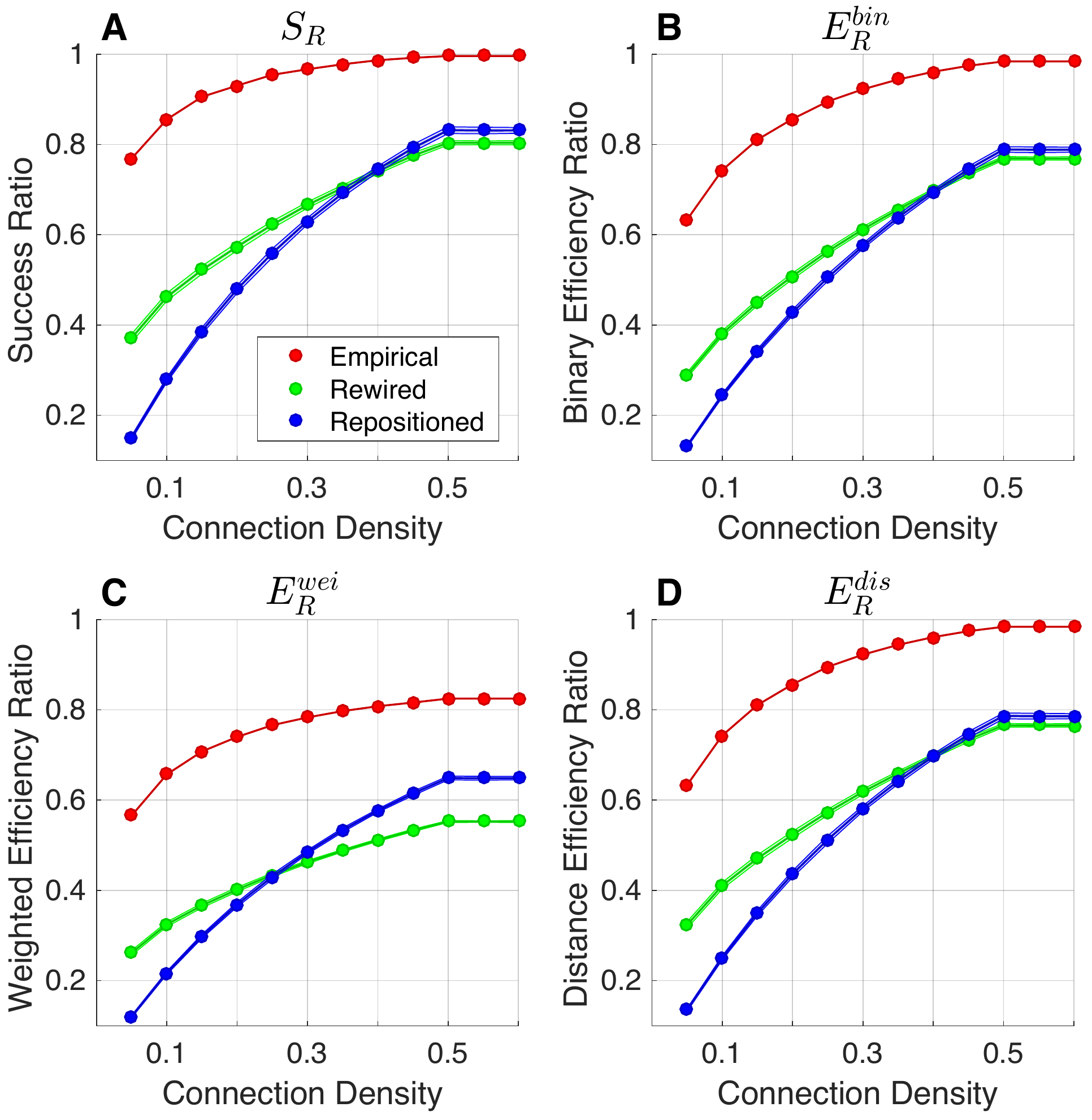}
			\label{fig:si_per_1024}
		}
		\caption{Navigation performance in human connectomes with different resolutions. \textbf{(A-D)} Success ratio ($S_{R}$), binary efficiency ratio ($E^{bin}_{R}$), weighted efficiency ratio ($E^{wei}_{R}$) and distance efficiency ratio ($E^{dis}_{R}$) for human structural networks at a $N=360$ parcellation resolution and different connection density thresholds. Empirical measures (red) for group-averaged connectomes were compared to 1000 ($N=256$), 500 ($N=512$) and 100 ($N=1024$) rewired (green) and spatially repositioned (blue) null networks. Shading indicates 95\% confidence intervals.}
		
		\label{fig:si_per}
		
\end{figure}
	
\begin{figure}[ht]	
		\centering
		\includegraphics[scale = 0.5]{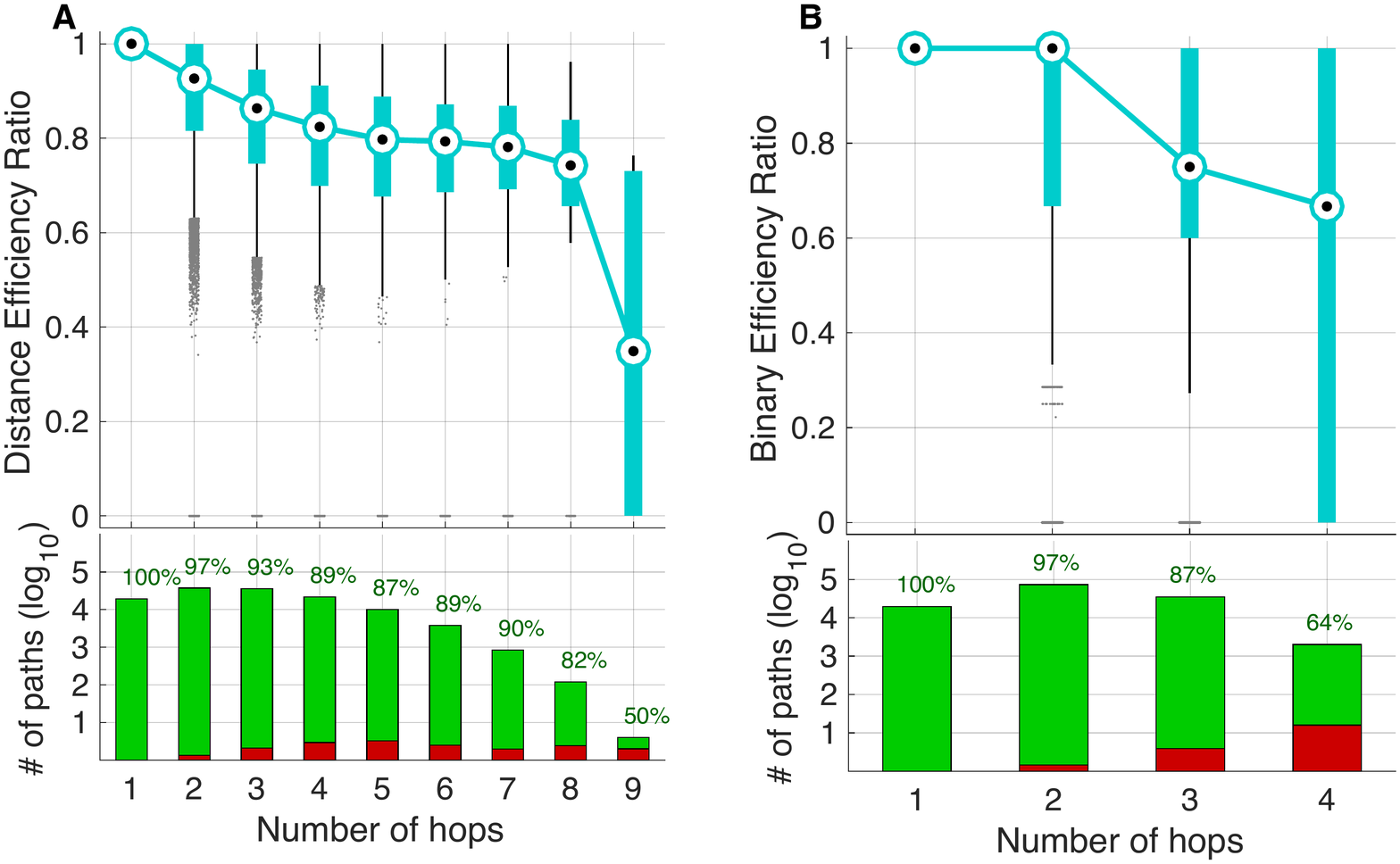}
		
		\caption{The performance of navigation routing in human SC stratified by shortest paths hop counts ($N=360$ at 15\% connection density). Blue boxplots indicate the quartiles of efficiency ratios of navigation paths benchmarked against shortest paths with matching hop count. Barplots show the number of shortest paths for a given hop count, with colors indicating the proportion of successful (green) and failed (red) navigation paths. Shortest paths computed over $L^{dis}$ \textbf{(A)} and $L^{bin}$ \textbf{(B)}.}
		
		\label{fig:si_mhp}
		
\end{figure}
	
%
%

\begin{figure}
	
	\centering
	\includegraphics[scale = 0.5]{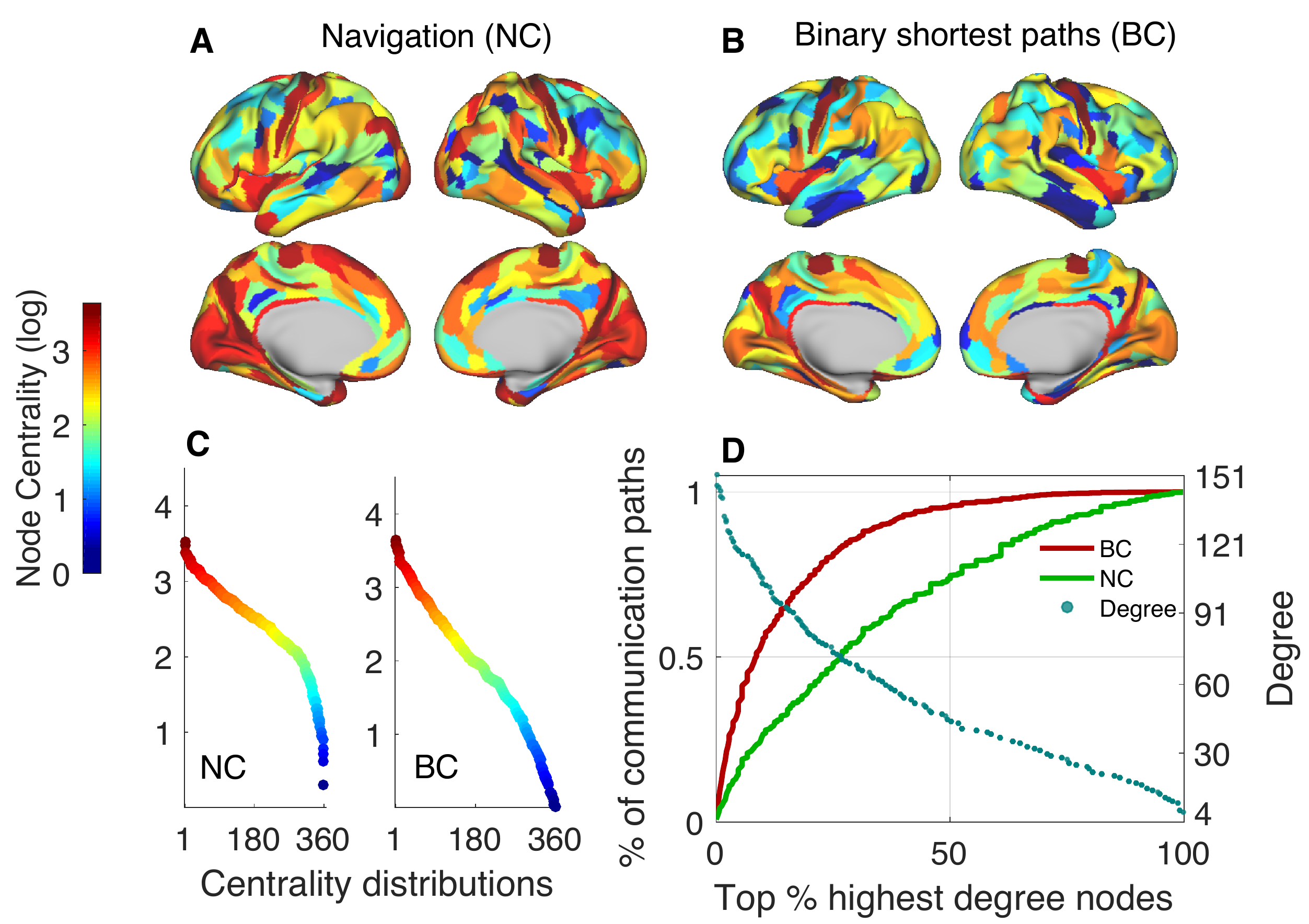}
	
	\caption{Comparison between navigation (NC) and binary betweenness (BC) node centralities for $N=360$ at 15\% connection density. Centrality values are logarithmically scaled. NC \textbf{(A)} and BC \textbf{(B)} projected onto the cortical surface. \textbf{(C)} NC (right) and BC (left) sorted from highest to lowest values \textbf{(D)} Relationship between the cumulative sum of centrality measures and degree. The horizontal axis is a percentage ranking of nodes from highest to lowest degree (e.g., for $N=360$, the 10\% most connected nodes are the 36 nodes with highest degree). Solid curves (left-hand vertical axis) represent the cumulative sum of BC (red) and NC (green) over all nodes ordered from most to least connected, divided by the total number of communication paths in the network, indicating the fraction of communication paths mediated by nodes. Blue dots (right-hand vertical axis) show the original degree associated with each percentage of most connected nodes.}
	
	\label{fig:si_cen}
	
\end{figure}

\begin{figure}[ht]	
	
	\centering
	\includegraphics[scale = 0.6]{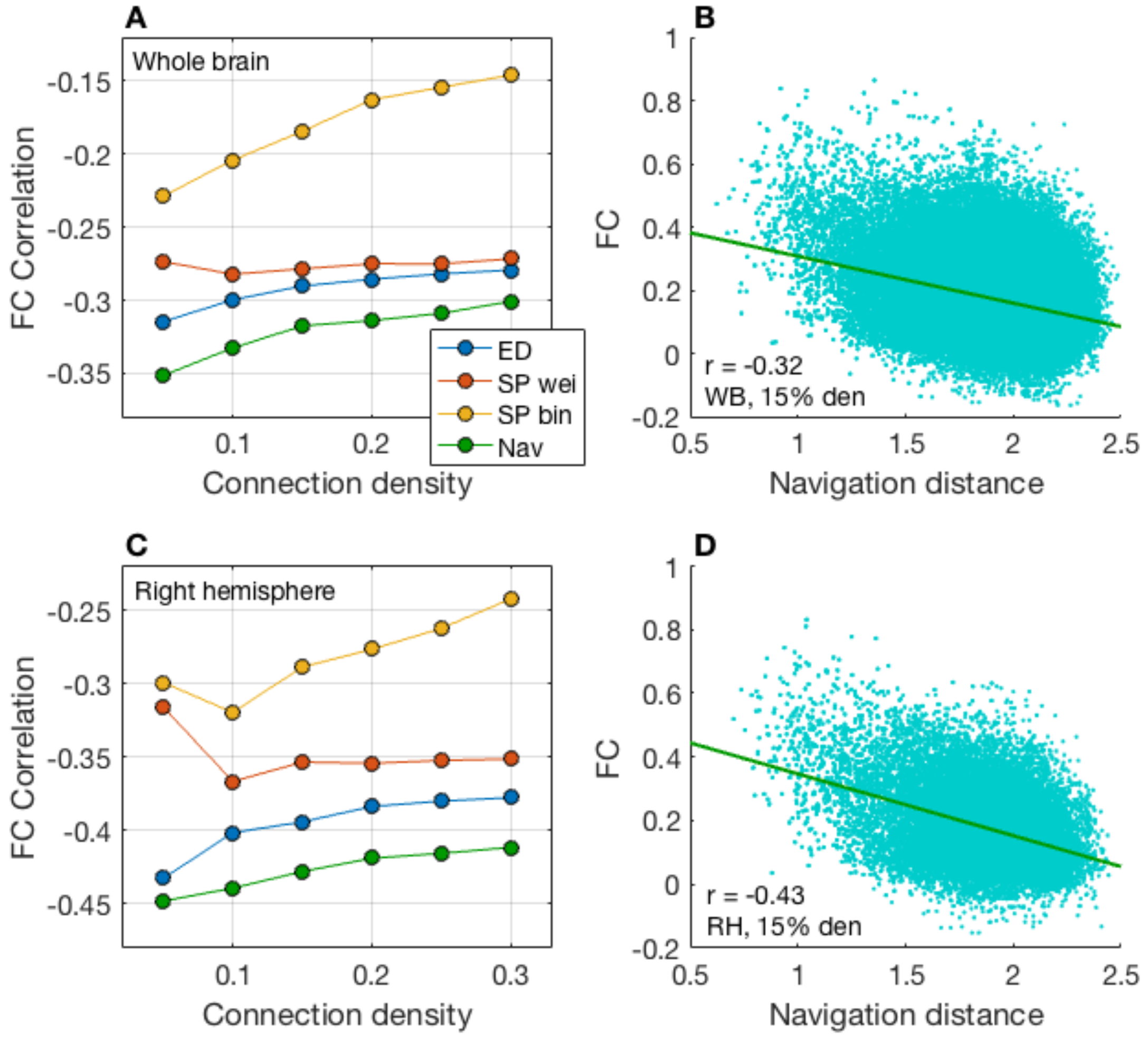}
	
	\caption{Correlation between FC and network communication measures ($N=360$, at 15\% connection density). Correlations were computed only for successful navigation paths. The navigation distance of nodes $i, j$ was defined as the log of the sum of connection distances along the navigation path from $i$ to $j$. \textbf{(A)} Whole brain correlations between FC and Euclidean distance (blue), weighted shortest paths (orange), binary shortest paths (yellow) and navigation distance (green). \textbf{(B)} Scatter plot between navigation distances and FC for all brain regions. Green line indicate the linear fit between the two measures. \textbf{(C-D)} Right hemisphere equivalent of panels A and D.}
	
	\label{fig:si_fcc}
	
\end{figure}
	
\begin{figure}[ht]	
	\centering
	\includegraphics[scale = 0.6]{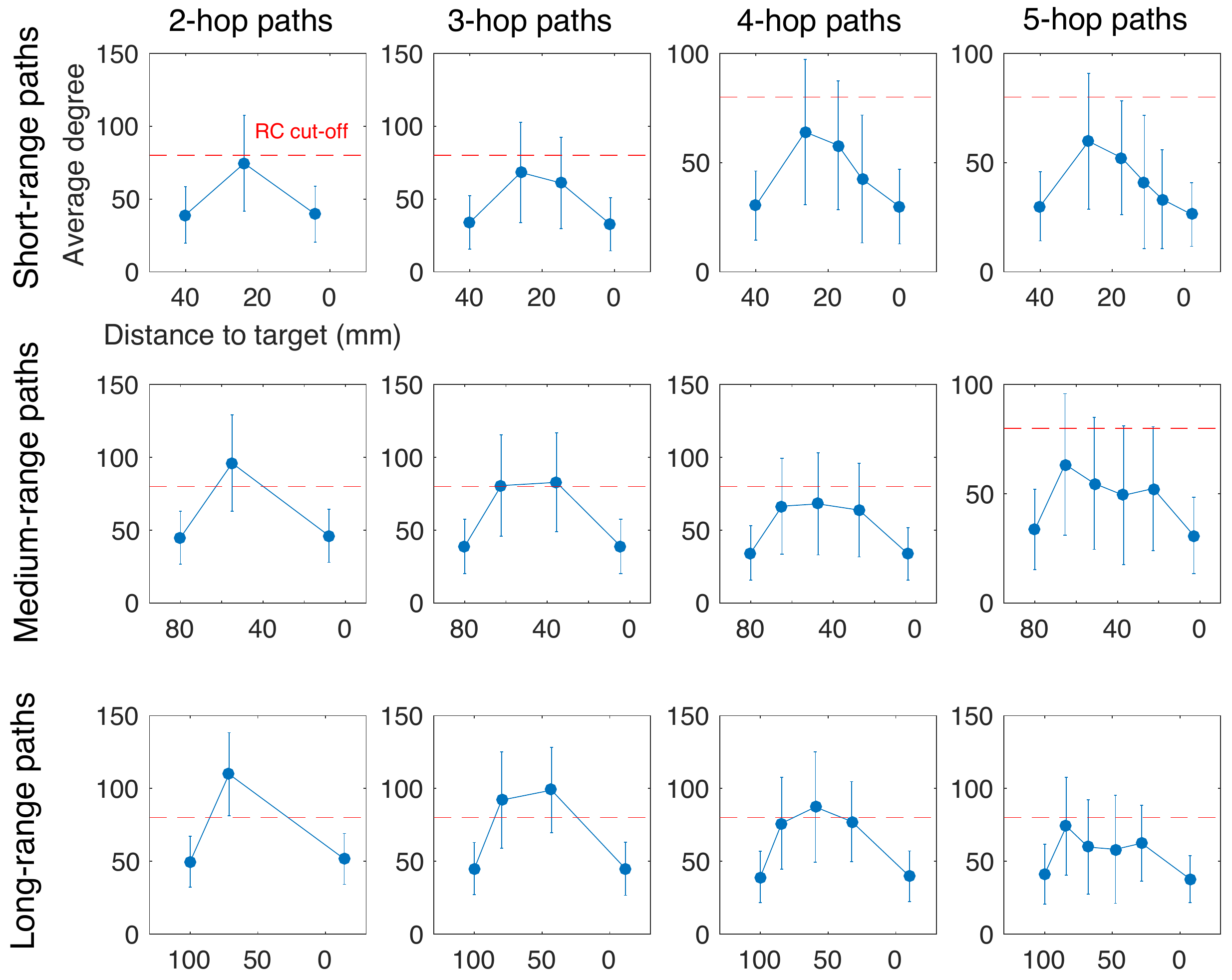}
	
	\caption{Structure of navigation paths between non rich club nodes ($N=360$ at 15\% connection density). Top, middle and bottom rows depict, respectively, short- (Euclidean distance from source to target nodes $d_{st}\leqslant$50 mm), medium- (50 mm$<d_{st}\leqslant$100 mm) and long-range paths ($d_{st}>$100 mm). Columns show paths 2 to 5 hops long. For each panel, the horizontal axis shows the average distance to the target, while the vertical axis shows the average degree (and standard deviation) of nodes along the path. The red dashed line marks the cut-off  degree $k=80$ for the binary rich club (computed as in \cite{heuvel:2012}). Navigation paths follow a ``zoom-out/zoom-in'' structure, proceeding allowing sequences of low-, high- and low-degree nodes. This path structure is facilitated by a heterogeneous degree distribution and high clustering coefficient. First, information travels via peripheral connections around the locally clustered vicinity of the source. Once it reaches a high-degree node, it is forwarded towards the target via long-range connections, quickly traversing large distances in a small number of hops. Finally, once at the destination's vicinity, the information again makes use of local clustering to home in on the target. This mechanism for information transfer is in line with the notion of the brain as a small world network that balances integration---high-degree nodes forming long-range connections that provide quick access to different parts of the brain---and segregation---specialized, locally clustered, information processing modules responsible for specific tasks. The rich club of highly interconnected hubs is predominantly used for long-range paths, while short- and medium-range communication takes place mostly via peripheral routes.}
	
	\label{fig:si_lhl}
	
\end{figure}

\clearpage

\bibliography{nav_bib_arxiv}
\bibliographystyle{ieeetr}